          [2001/04/25 1.1 (PWD)]
\documentclass[a4paper,twocolumn]{esapub2005} 
\usepackage{times}
\usepackage{natbib}
\usepackage{graphicx}

\title{Population of HMXB in the Galaxy}
\author[1]{Alexander A. Lutovinov}
\author[2,1]{Mikhail G. Revnivtsev}
\author[2,1]{Marat R. Gilfanov}
\author[2,1]{Rashid A. Sunyaev}
\affil[1]{Spase Research Institute, Profsoyuznaya str. 84/32,
Moscow  117997, Russia}
\affil[2]{Max-Planck-Institut f\"ur Astrophysik, Karl-
Schwarzshchild str. 1, Garching, Germany}

\begin{document}
\renewcommand{\topfraction}{1.0}
\renewcommand{\bottomfraction}{1.0}
\renewcommand{\textfraction}{0.0}


\maketitle

\begin{abstract}

We study populations of High-Mass X-ray Binaries in the Galaxy using
data of the INTEGRAL observatory in a hard X-ray energy band. More
than two hundreds of sources were detected with INTEGRAL near the
galactic plane ($|b|<5$ deg), most of them have a galactic origin and
belong to high (HMXB) and low mass (LMXB) X-ray binaries. We
investigated properties and spectra of a large sample of HMXBs and
concluded that most of them are belong to X-ray pulsars. We also build
the distribution of HMXBs for the whole Galaxy and showed that its
peaks are practically coincident with spiral arm tangents. The
obtained results are discussed in terms of some model estimations of
the density of different components of the Galaxy.

\end{abstract}

\section{Introduction and previous works}

The INTEGRAL observatory is successfully operating on the orbit about
$~3.5$ years since its launch in 2002 \citep{wink03}. For this period
more than 35 million seconds of data were provided and analyzed. One
of the most interesting results obtained by INTEGRAL for this period
are surveys of different parts of the sky and especially the Galactic
center and plane (\citep{rev04,mol04,rev06,bird06}). These surveys as
well as other INTEGRAL observations allowed to discover several dozens
sources of different nature, mostly galactic X-ray binaries. These
objects are brightest X-ray emitters in the Galaxy and their positions
in the Galaxy connected with the origin of their optical star.  Based
on the RXTE/ASM sky survey in the $2-10$ keV energy band Grimm et
al. \citep{gri02} showed that the spatial distribution of high mass
X-ray binaries (HMXB) and low mass X-ray binaries (LMXBs) is
significantly different. HMXBs being the young X-ray population of the
Galaxy should trace the star formation (SF) regions, while LMXBs
should be more concentrated in the regions of high stellar mass
density, particularly in the Galactic center region (see also
\citep{gre96}).

INTEGRAL observations and discoveries of a large number of X-ray
sources in a hard energy range ($>20$ keV) where the photoabsorption
doesn't play any role allowed to significantly move in the study of
different populations of X-ray binaries, their properties and
spatial distribution. In our first paper devoted to this problem
\citep{lut05a} we were focused on a sample of Galactic sources
located in the Galactic plane between the Norma and Sagittarius
spiral arms (the inner part of the Galaxy). The following
investigations used more than $24$ Msec of publicly available data
of INTEGRAL observations and were extended to the whole Galactic
plane \citep{lut06}.

In frame of work dedicated to the all-sky survey in hard X-rays
\citep{kris07} we analyzed all available at the moment data of the
INTEGRAL observatory and studied the population of X-ray binaries in
the Galaxy. Some preliminary results concerned HMXBs are briefly
presented in the current paper.

\section{Individual sources, their properties and spectra}

There is a large population of highly photoabsorbed Galactic X-ray
sources among newly discovered ones with the INTEGRAL observatory. The
prominent feature of such sources is a strong intrinsic absorption
($\sim10^{23}-10^{24}$ cm$^{-2}$), that made difficult their
detection in a soft energy band ($<10$ keV). Remarkably these sources
were confined in relatively small regions of the sky, close to the
tangents of the Galactic spiral arms (e.g. \citep{lut05a,lut06}). The
analysis of their properties showed that they are high mass X-ray
binaries with early type companion stars (giant or supergiant) with a
strong stellar wind (e.g. \citep{rev03}). The usage of hard X-ray
energy band $>20$ keV (ISGRI detector of the INTEGRAL/IBIS telescope,
\citep{leb03}) allowed one to reconstruct spectra of such heavily
absorbed sources and showed that most of them are very similar to spectra
of X-ray pulsars and can be fitted the powerlaw model with a
cutoff(\citep{lut05a}). The following studies of these sources
performed with XMM-Newton, RXTE, Chandra observatories revealed
coherent pulsations from some of them (e.g. \citep{pat04,wal04,lut05b}).

\begin{figure*}[t]
\vbox{
\hbox{
\includegraphics[width=0.5\textwidth,bb=30 420 575 740]{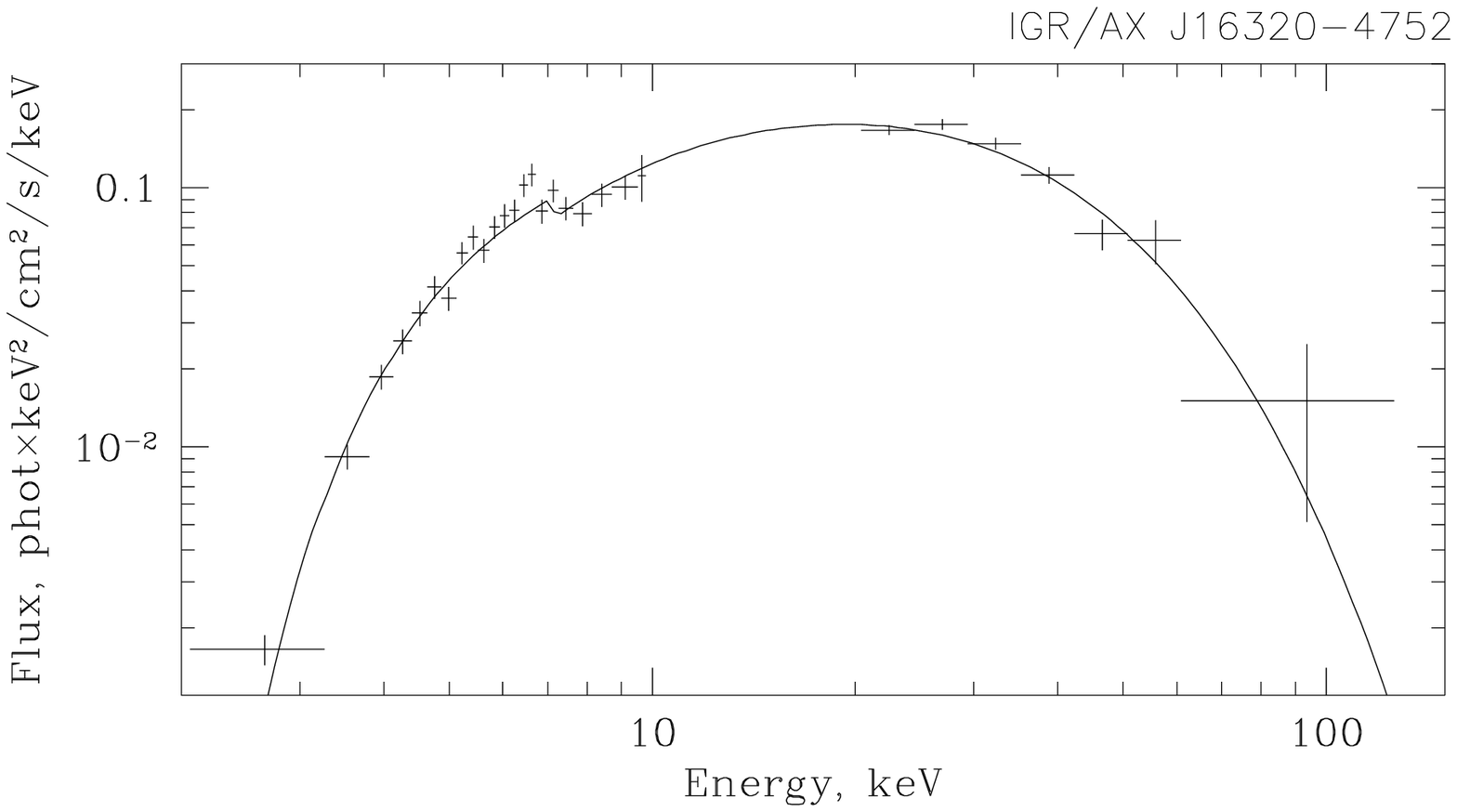}
\includegraphics[width=0.5\textwidth,bb=30 420 575 740]{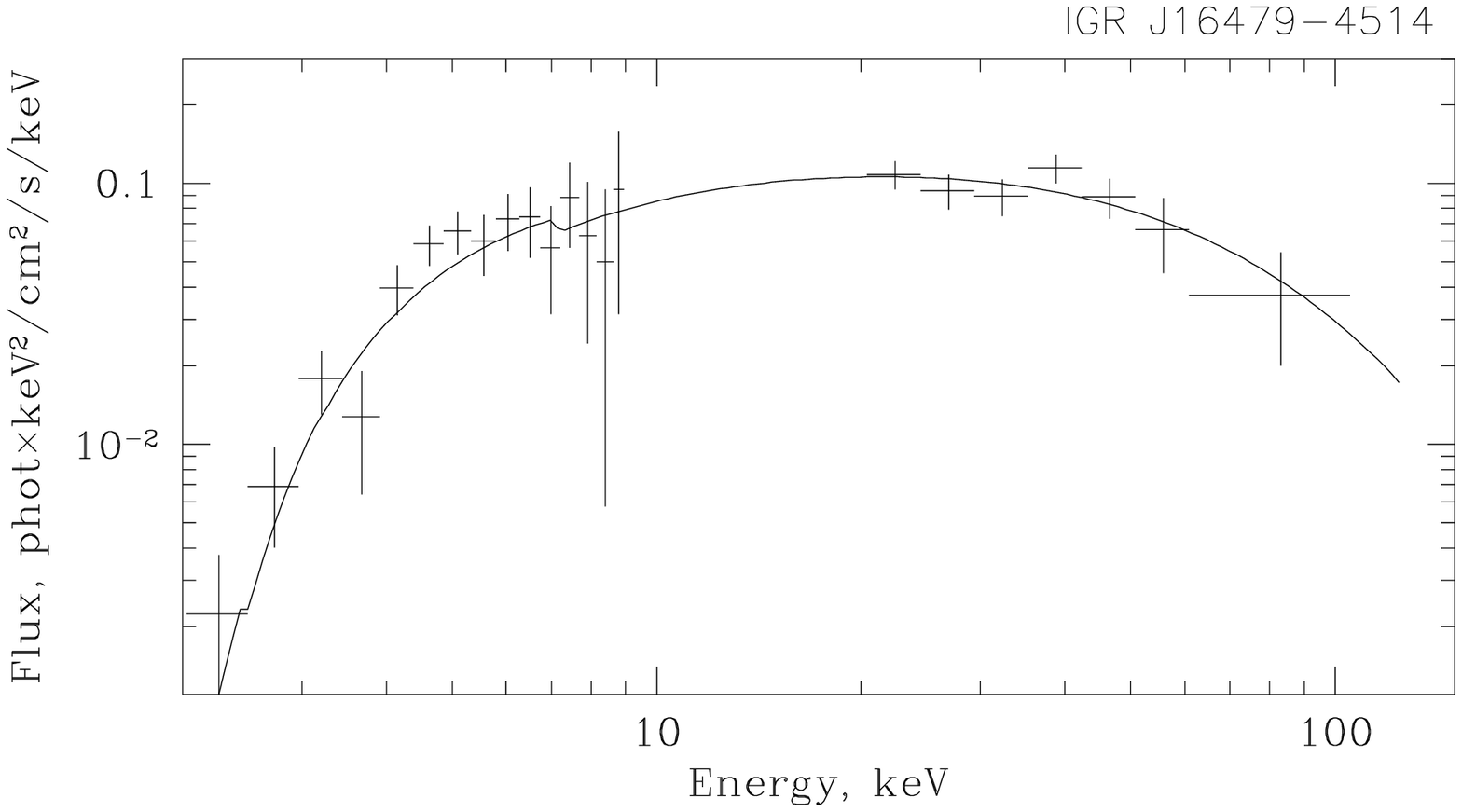}
} 
\includegraphics[width=1.02\textwidth,bb=88 310 570 560]{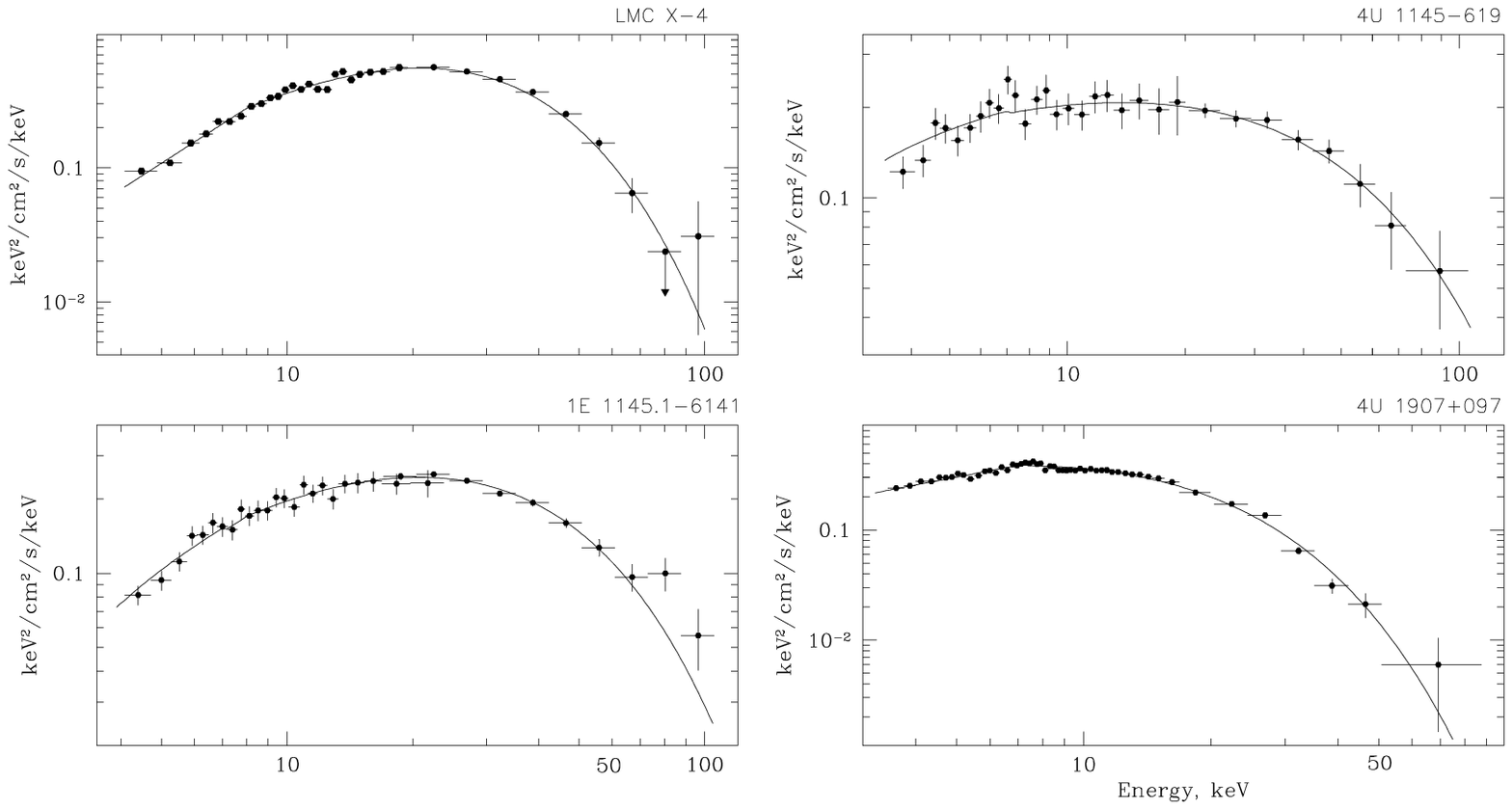}
}
\caption{({\it Two upper panels}) Broadband energy spectra of two new 
absorbed sources IGR/AX J16320-4752 and IGR J16479-4514, obtained by
XMM-Newton and INTEGRAL observatories. ({\it Bottom panels}) Typical
broadband spectra of four known X-ray pulsars obtained by the JEM-X
monitor and IBIS telescope of the INTEGRAL observatory (see
\citep{fil05} for details). The best-fit models are shown by solid
lines.}\label{spc_abs}
\end{figure*}

As an example in Fig.\ref{spc_abs} spectra of new heavily absorbed
sources IGR/AX J16320-4752 and IGR J16479-4514 obtained with INTEGRAL
($>20$ keV) and XMM-Newton ($1-10$ keV) observatories are
presented. The spectra of 4 known X-ray pulsars obtained in a wide
energy band (4-100 keV) with the JEM-X and IBIS telescopes of the
INTEGRAL observatory are shown in Fig.\ref{spc_abs} for the
comparison. It is clearly seen that spectra of INTEGRAL sources and
X-ray pulsars are very similar. The subsequent study with using data
of the XMM-Newton observatory and archival data of the ASCA
observatory actually revealed pulsations from IGR/AX J16320-4752 with
the period of $\sim1300$ s (\citep{lut05b}). Thus X-ray pulsars
constitute a significant part of high-mass X-ray binaries as among
newly discovered sources as among those observed with INTEGRAL. In
particular, most of HMXBs observed with the INTEGRAL observatory in
the inner part of the Galaxy (16 of 23) are accretion-powered X-ray
pulsars (\citep{lut05a}). In total the INTEGRAL observatory
significantly detected and allowed to reconstruct spectra of 35 X-ray
pulsars (\citep{fil05}).

Note, that despite of the general similarity of spectra of new sources
and X-ray pulsars the more strong absorption in new sources is
obviously seen (Fig.\ref{spc_abs}).

\section{Spatial distribution of X-ray binaries}

\begin{figure*}[t]
\includegraphics[width=\textwidth,bb=0 0 610 150]{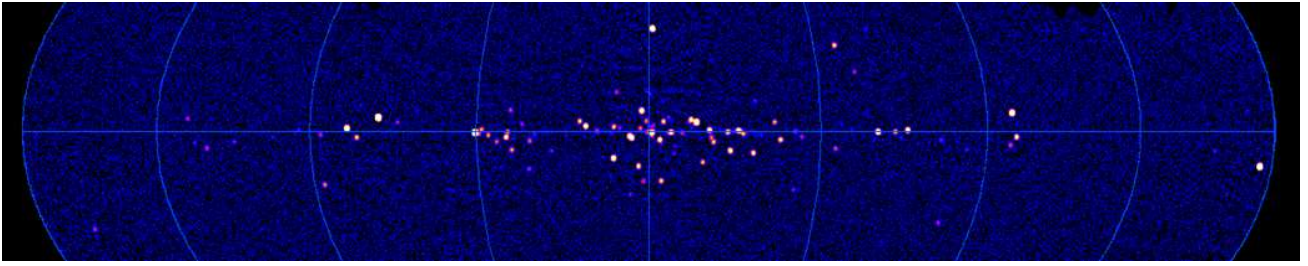}
\caption{Map of the sky region near the Galactic plane obtained with 
ISGRI/IBIS/INTEGRAL in the $17-60$ keV energy band.}\label{gal_plane}
\end{figure*}

At the moment (June of 2006) there are more than $35$ Msec of publicly
available data of INTEGRAL observations. These observations allow us
to perform the all-sky survey in hard X-rays (\citep{kris07}) and
study both extragalactic (\citep{saz06}) and Galatic sources. About
400 sources were detected in all data with a high statistical
significance; about 140 from them are new sources discovered by
INTEGRAL. As is known X-ray binaries are concentrated towards the
Galactic plane, however HMXBs and LMXBs have different vertical scale
heights, reflecting the age of stellar companions of these sources:
$\sim150$ pc for HMXBs and $\sim400$ pc for LMXBs (\citep{gri02}). At
the galactic Center distance from the Sun (assume 8.5 kpc) these scale
heights correspond to angular scales of $\sim1$ deg and $\sim2.7$ deg,
correspondingly. The part of the sky near the Galactic plane observed
with the INTEGRAL observatory in the $17-60$ energy band is shown in
Fig.\ref{gal_plane}. Following to \citep{lut06} below we present two
current samples of sources with $b<|2|$ deg and $b<|5|$ deg from the
Galactic plane.

\begin{center}
\begin{tabular}{lc}

\hline\\[-2mm]
\multicolumn{2}{c}{All sky}\\[1mm]
\hline\\[-2mm]
Total & 396 sources\\ HMXB & 60 sources\\ LMXB & 76 sources\\[1mm]
\hline\\[-2mm]
\multicolumn{2}{c}{Galactic plane ($b<|5|$, LMXB scale)}\\[1mm]
\hline\\[-2mm]
Total & 228 sources\\
HMXB & 50 sources \\
LMXB  & 55 sources\\[1mm]
\hline\\[-2mm]
\multicolumn{2}{c}{Galactic plane ($b<|2|$, HMXB scale)}\\[1mm]
\hline\\[-2mm]
Total  & 143 sources\\
HMXB & 41 sources \\
LMXB & 33 sources\\[1mm]
\hline
\end{tabular}
\end{center}

\begin{figure}[b]
\hspace{-2mm}\includegraphics[width=0.49\textwidth,bb=40 200 570 680]{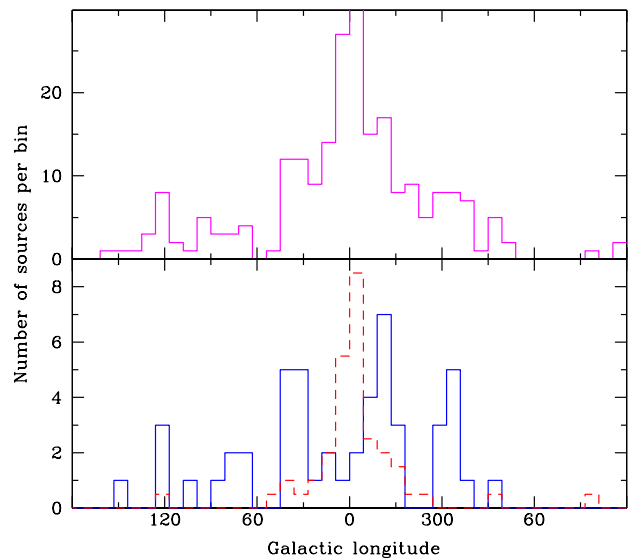}
\caption{Angular distributions of all detected sources (upper panel), 
identified HMXBs and LMXBs (solid and dashed lines in bottom panel,
respectively). The number of LMXBs are divided by 2.}\label{hist}
\end{figure}

It is interesting to note that relative numbers of HMXBs and LMXBs
changes considerably if we widen our selection region with the respect
to the Galactic plane, that reflects different vertical scales of
distributions both type of sources. Note, that majority of absorbed
sources, discovered by INTEGRAL, lies very close to the Galactic
plane. In Fig.\ref{gal_plane} we can clearly see a large number of
sources near the Galactic plane and rapid drop of their surface
density towards higher Galactic latitues. Another interesting point is
that only 3 black hole candidates and 32 X-ray pulsars detected among
50 HMXBs inside of the 5-deg region from the Galactic plane; if we go
down to the 2-deg region only 1 black hole candidate and 23 X-ray
pulsars will be indentified from 41 HMXBs in total.

\begin{figure*}[t]
\hbox{
\includegraphics[width=0.47\textwidth]{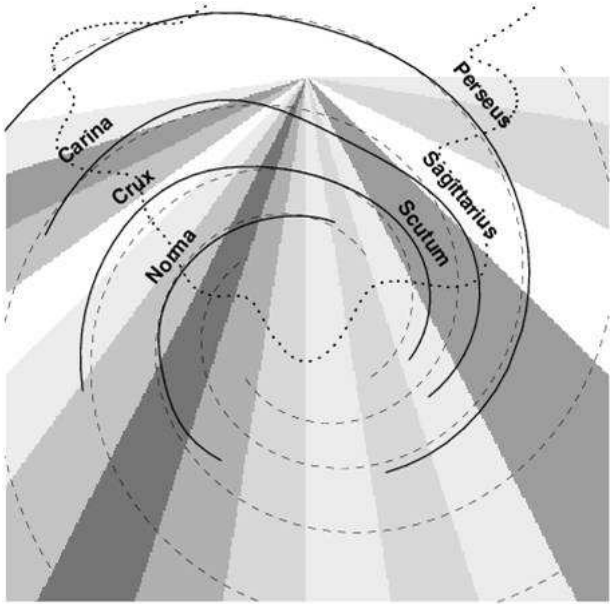}
\hspace{4mm}\includegraphics[width=0.48\textwidth]{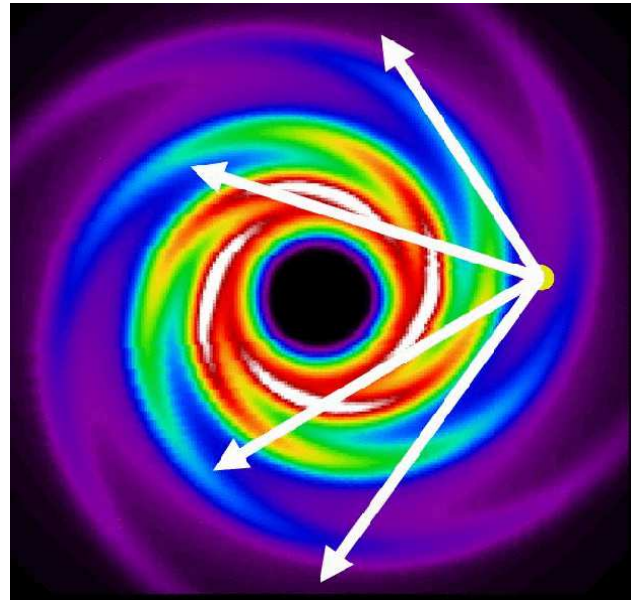}
}
\caption{({\it Left}) Face-on view of the Galaxy with overlayed densities 
of HMXBs shown in gray scale. Solid lines show the position of the
spiral arms by \citep{taylor93}; the dashed line shows logarithmic
spiral arms by \citep{val95}; the dotted line represents the current
sensitivity of the survey at the level of $L_{\rm x}=10^{35}$
erg/s. ({\it Right}) The model mass distribution in the Galactic disk
with the superimposed spiral structure. White arrows show the
directions of peaks of the HMXBs concentration.}\label{gal_struct}
\end{figure*}

Our Galaxy consists of four components -- disk, bulge, spheroid and
halo, the former two bring the main contribution to the mass of the
Galaxy (e.g. \citep{bah86}). Moreover, there is a clear spiral
structure in the Galactic disk which is believed to be a spiral
density wave, initiating the intense star formation. Therefore we
might naturally expect that increased number density of high mass
X-ray binaries should be observed in spiral arms regions. In
Fig.\ref{hist} the distribution of the surface density of HMXB along
the Galactic plane is presented in comparison with the LMXB one. Note,
that samples of sources considered here are not flux limited therefore
ununiformity of exposure times should be take into account in the
observed sources distributions. It is obviously seen that there are
concentrations of HMXB in the regions of tangents to the spiral arms,
while it is much weaker in the LMXBs distribution, that have a maximum
near the Galactic center. To estimate the significance of this
difference quntitavely we used a Kolmogorov-Smirnov test. We build
cumulative distributions for each of samples and found that
probabilities that the HMXB distribution differs from LMXB and uniform
ones are $>99.9$\% in both cases.

For the better visualization the observed distribution of HMXBs in the
Galaxy is presented in Fig.~\ref{gal_struct} (left panel) with respect
to the Galactic spiral structure. The number density of the sources is
shown in gray scale; the spiral model of the Galaxy is based on
optical and radio observations of HII regions (\citep{taylor93}) and
logarithm spirals with pitch angle $12^\circ$ adopted by
(\citep{val95}). The dotted line shows maximal distances from the Sun
at which the sensitivity of the IBIS telescope and current exposure
allow us to detect X-ray sources with the luminosity $L_{\rm
x}=10^{35}$ erg/s. Again, it is clearly seen that the HMXBs
distribution strictly avoids the Galactic center and is concentrated
towards Galactic spiral arms. Such result can be understand from the
current knowledge of the Galactic structure. In Fig.~\ref{gal_struct}
(right panel) the face-on distribution of the mass in the Galactic
disk with the superimposed spiral arms structure is presented. The
white arrows show the believed directions of the enhanced number of
high-mass X-ray binaries, that are very close to the observed ones.
Note, that despite of understanding of the HMXBs distribution in
general, some subtle details are not clear yet. In particular, there is
no exact coincidence of distribution peaks and spiral arm tangents,
that was mentioned firstly for the inner part of Galaxy by
(\citep{lut05a}) and confirmed here for the whole Galaxy.

\section*{Acknowledgments}

We would like thank to E.Churazov for developing methods of analysis
of the IBIS data and software. This work was supported by the Russian
Academy of Sciences (The origin and evolution of stars and galaxies
program), grant of President of RF (NSh-1100.2006.2) and grant of RFBR
07-02-01051. AL acknowledges the financial support from the Russian
Science Support Foundation.

\end{document}